# An Integrated Biological Optimization framework for proton SBRT FLASH treatment planning allows dose, dose rate, and LET optimization using patient-specific ridge filters

Ruirui Liu[1], Serdar Charyyev[1], Niklas Wahl[2,5], Wei Liu[3], Minglei Kang[4], Jun Zhou[1], Xiaofeng Yang[1], Filipa Baltazar[2,6], Martina Palkowitsch[2,5,7], Kristin Higgins[1], William Dynan[1], Jeffrey Bradley[1], Liyong Lin[1]

[1]Department of Radiation Oncology and Winship Cancer Institute, Emory University, Atlanta, GA 30322, USA

[2]German Cancer Research Center – DKFZ, Department of Medical Physics in Radiation Oncology, Heidelberg, Germany

[3]Department of Radiation Oncology, Mayo Clinic, Phoenix, Arizona, USA

[4]New York Proton Center, New York, New York, USA

[5]Heidelberg Institute for Radiation Oncology – HIRO, Heidelberg, Germany

[6]Instituto SuperiorTécnico, Universidade de Lisboa, Av. Rovisco Pais,1, 1049-001 Lisbon, Portugal

[7]Atominstitut, TU Wien, Stadionallee 2, 1020 Vienna, Austria

## Abstract

**Purpose:** Patient-specific ridge filters provide a passive means to modulate proton energy to obtain a conformal dose. Here we describe a new framework for optimization of filter design and spot maps to meet the unique demands of ultra-high dose rate (FLASH) radiotherapy. We demonstrate an Integrated Biological Optimization IMPT (IBO-IMPT) approach for optimization of dose, dose-averaged dose rate (DADR), and dose-averaged LET (LETd).

**Methods**: We developed inverse planning software to design patient-specific ridge filters that spread the Bragg peak from a fixed-energy, 250 MeV beam to a proximal beam-specific planning target volume (BSPTV). The software defines patient-specific ridge filter pin shapes and uses a Monte Carlo calculation engine, based on Geant4, to provide dose and LET influence matrices. Plan optimization, using matRAD, accommodates the IBO-IMPT objective function considering dose, dose rate, and LET simultaneously with minimum MU constraints. The framework enables design of both regularly spaced and sparse-optimized ridge filters, from which some pins are omitted to allow faster delivery and selective LET optimization. Volume distributions and histograms for dose, DADR, and LETd are compared using evaluation structures specific to the heart and lung.

**Results**: To demonstrate the framework, we used IBO-IMPT to design ridge filters for a central lung tumor patient. The IBO-IMPT framework selectively spared heart and lung by reducing LET and increasing dose rate, relative to conventional IMPT planning. Sparse-optimized ridge filters were superior to regularly spaced ridge filters in dose rate. Together, these innovations substantially increased the DADR in the heart and lung while maintaining good dose coverage within the BSPTV. The volume that received a FLASH dose rate of ≥ 40 Gy/second increased by 31% for heart and 50% for lung.

**Conclusion**: This proof-of-concept study demonstrates the feasibility of using an IBO-IMPT framework to accomplish proton FLASH SBPT, accounting for dose, DADR, and $LET_d$ simultaneously.

2

# Introduction

Although stereotactic body radiation therapy (SBRT) provides excellent local tumor control, it poses unacceptable risks in a subset of patients. For example, patients with central and ultra-central lung tumors are at a 15% risk of fatal hemorrhage based on impingement of the complex overlapping radiation fields on organs at risk (OARs), including uninvolved lung, heart, and esophagus [1–4]. Stereotactic body proton therapy (SBPT) represents an advancement over SBRT [5,6], as it uses fewer beams and delivers much of the dose in a patient-specific spread-out Bragg Peak (SOBP), sparing proximal and especially distal OARs. Even with SBPT [6,7], there is necessarily some treatment margin [8–12], which may impact OARs and thus limit clinical applicability [13].

FLASH radiotherapy is a novel modality with the potential to provide further sparing of OARs beyond that offered by conventional SBPT. Recent mouse studies using electron or scattered proton transmission FLASH suggest that the toxicities to serial OARs, such as the esophagus, may be reduced by 30%-50%, while maintaining anti-tumor efficacy[14,15]. The current generation of proton therapy machines are, in many cases, capable of achieving FLASH dose rates (e.g., 40-800 Gy/second). Typically, irradiation is performed using a high-energy transmission beam[16]. Energy modulation is impractical, given that characteristic energy modulation times (>500 milliseconds)[17–19] exceed the total time allowed for FLASH delivery (250 milliseconds for a typical 10 Gy SBPT dose). Unfortunately, the use of the transmission beam sacrifices a major advantage of proton therapy: the ability to deliver dose in a SOBP. For small SBPT targets other than extremities, the increased spillover to serial OARs can more than offset FLASH sparing.

There are several passive, and thus dose rate-independent, approaches that can be used to improve conformality at FLASH dose rates[20–23]. One is to use a range compensator to position a Bragg peak within the target. For a high beam-current proton facility, multi-field optimized IMPT with a universal range shifter and compensator can achieve FLASH delivery[22]. Another approach is to use a universal machine-specific ridge filter in scattered proton mode[15,24,25]. This achieves a SOBP with a single field, although the dose distribution is not conformal to the target. A third approach, which is the subject of the the present study, is to combine a patient-specific ridge filter with a range compensator to achieve a single field-optimized (SFO), IMPT-like conformal dose distribution[18,26].

Although patient-specific ridge filters have clear advantages, designing such filters is challenging. Specifically, there is an unmet need to optimize ridge filter design and spot maps to maximize sparing of organs at risk. Because dose, dose-averaged dose rate [27], and dose-averaged LET ($LET_d$)[28,29] each influence the biological response, simultaneous optimization of all three factors is desirable. The integrated biological optimization (IBO)-IMPT framework described here achieves this goal and thus represents an advance over other currently available approaches. Using this framework, we found that sparse ridge filters (from which some pins have been omitted) offer advantages over regularly-spaced ridge filters (hereafter referred to as regular ridge filters) by enabling higher dose rates at specific locations, maximizing the FLASH effect.

## Materials and Methods

### 3D Ridge filter design

The beam-specific planning target volume (BSPTV) [8,30] is used to design the patient-specific ridge filters. Fig. 1 illustrates the methodology for filter design and fabrication. Fig. 1A shows an enlarged view of a stepwise-tapered single ridge pin. Each modulation step creates a separate Bragg peak. The weight (cross sectional area of the step) and thickness (height) of each step are denoted by variables $w_i$ and $t_i$. These are used as inputs to the equation (1), where $D_i$ represents the dose at the $i$-th Bragg peak and $B(t_i, j)$ is the depth dose of the Bragg peak by the $i$-th modulation ridge thickness at position $j$. We generate the $B(n,m)$ matrix of $n$ Bragg curves consisting of $m$ points through Geant4 simulation. We generate the ridge filter information file by solving the equation set (1) using the least square method (equation (2)) to provide the area $w_i$ of thickness $t_i$.

$$\begin{cases} D_1 = B(t_1,1)w_1 + B(t_2,1)w_2 + \cdots B(t_N,1)w_N \\ D_2 = B(t_1,2)w_1 + B(t_2,2)w_2 + \cdots B(t_N,2)w_N \\ \vdots \\ D_M = B(t_1,M)w_1 + B(t_2,M)w_2 + \cdots B(t_N,M)w_N \end{cases} \quad (1)$$

$$\operatorname*{argmin}_{w} \frac{1}{2}\|Bw - D\|^2 \quad (2)$$

Achieving the desired, IMPT-like dose distribution requires filter pin spacing that is much smaller than the Gaussian sigma of the proton spots. This assures that proton energies will be mixed in the desired proportion.

We then extend the single pin optimization to multiple pins, which are arranged to cover the whole tumor volume. The optimized weight factors are translated to the geometrical parameters of the filter pin. The filter pin positions are defined in beam eye view (BEV) (Fig. 1B). The complete assembly includes both filter pins and a range compensator (Fig. 1C). FreeCAD (http://www.freecadweb.org) is then used to generate a 3D printing stereolithography file, and the filter assembly is printed using Accura Extreme[31], which has a density of about 1.19 g/cm$^3$ (Fig. 1D). Micro-CT image is used to demonstrate that the ridge filter conforms to the design (Fig. 1E). We developed an in-house full Monte Carlo dose calculation engine for the patient-specific ridge filter using Geant4 (Version 10.7) to calculate the patient dose with a ridge filter. A radiation transport schematic is shown in Fig. 1F.

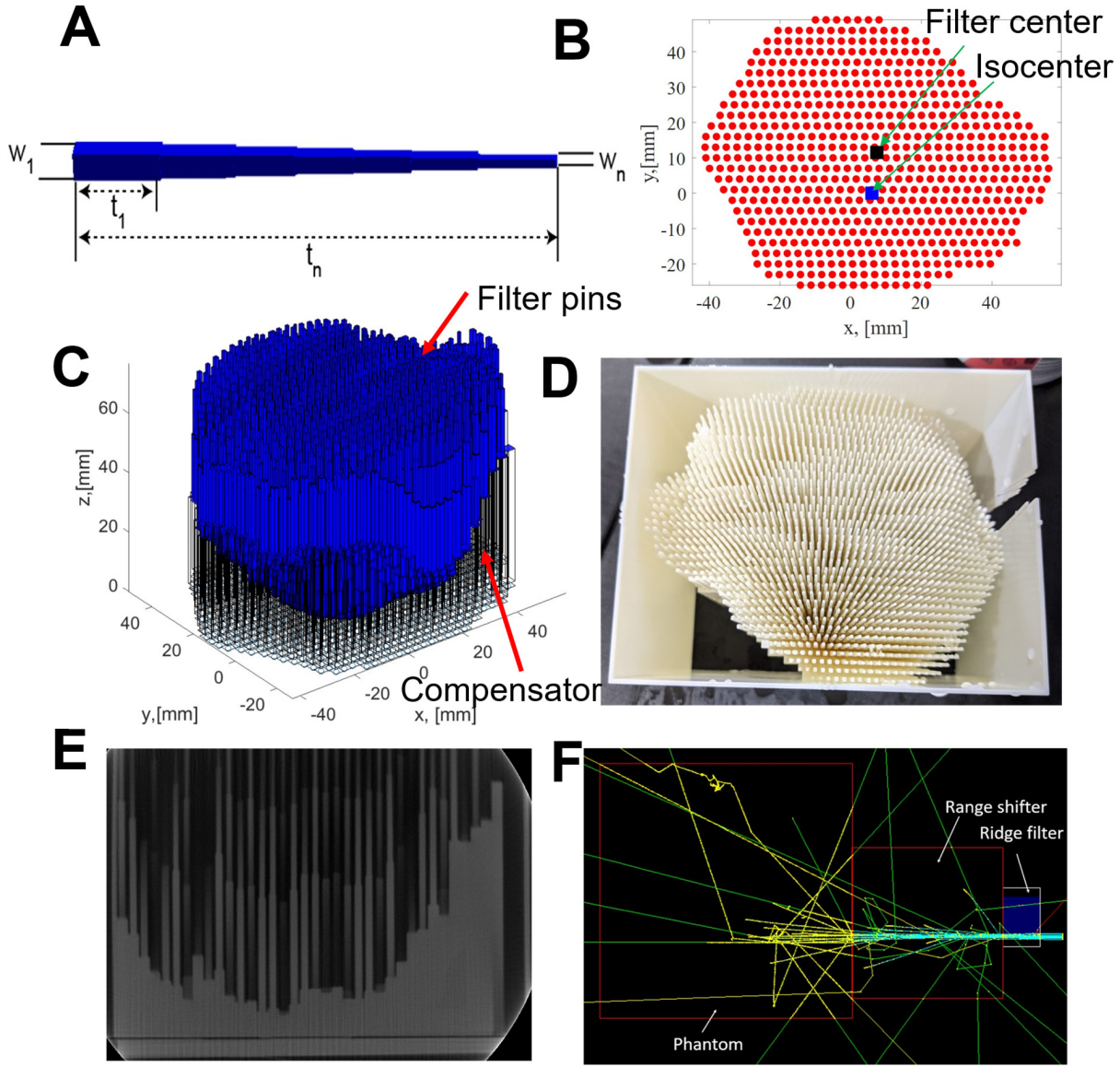

Figure 1: Design and fabrication of 3D printed ridge filters. A. Single ziggurat-shaped filter pin that is optimized to provide a 4 cm SOBP for a 250 MeV proton beam. B. Example filter pin location map for a lung tumor target. Red dots indicate the filter pin locations of the ridge filter in BEV coordinates. The geometrical center of the ridge filter and the treatment plan isocenter are marked. C. Complete assembly consisting of filter pins and range compensator. D. Example of a 3D printed ridge filter. E. Micro-CT scan of the printed ridge filter showing one slice in coronal view (Precision X-ray Inc., North Branford, CT). The field of view is 5 cm × 5 cm with a spatial resolution of 50 μm. F. Full Monte Carlo radiation transport simulation schematic for a ridge filter.

**Treatment planning system**

The patient's 3D voxelized geometry file, ridge filter information file, and beam parameters including gantry angle and initial spot map are fed into Geant4 to obtain the 3D dose and $LET_d$ influence matrices [32,33] that serve as inputs for inverse optimization of spot weights, as shown in Fig. S1 in Supplementary Materials. The open-source treatment planning toolkit, matRad [34,35], was used to develop a treatment planning system (TPS), implementing the IBO-IMPT framework to generate an optimized spot map that conforms to the target dose coverage and OAR constraints specified in the treatment plan. The matRad-based TPS determines the optimized spot map to meet the minimum MU constraint[16]. MatRad is written in MATLAB and relies on an interior point optimization package (IPOPT) [36] to solve the fluence optimization problem. It uses L-BFGS with a logarithmic barrier[34] to implement the required boundary constraints. Specifically, we optimize the spot map based on the equation (3), which is the main objective function for optimization. A DADR quadratic term has been added to the objective function originally proposed by Liu and coworkers[32,33], to allow simultaneous optimization of dose, DADR, and $LET_d$ simultaneously. Equation (4) describes the dose summation process using the weighted dose influence matrix. Equation (5) describes the calculation of DADR. Equation (6) describes the calculation of the minimum MU [37].

$$\min f(w) = \sum_{t \in T} \frac{\alpha_t}{N_t} \sum_{i=1}^{N_t} \left( \max\{0, \eta_t - d_i(w)\} \right)^2 + \sum_{o \in OAR} \frac{\alpha_o}{N_o} \sum_{i=1}^{N_o} \left( \max\{0, d_i(w) - \eta_o\} \right)^2$$
$$+ \sum_{o \in OAR} \frac{\alpha_{o,DADR}}{N_o} \sum_{i=1}^{N_o} \left( \max\{0, DADR_o - DADR_i(w)\} \right)^2 \quad (3)$$
$$+ \sum_{o \in OAR} \frac{\alpha_{o,LET}}{N_o} \sum_{i=1}^{N_o} \left( \max\{0, LETd_i(w) - LETd_o\} \right)^2$$

$$d_i = \sum_j D_{i,j} w_j \quad (4)$$

$$DADR_i = \frac{1}{D_i} \sum_{j=1}^{N_o} (D_{ij} W_j)(D_{ij} I_j) \quad (5)$$

$$W_{j,\min} = \frac{I_{nozzle} \times T_{\min}}{N_{MU}} \quad (6)$$

$$LET_{di} = \frac{1}{D_i} \sum_{j=1}^{N_0} L_{ij} D_{ij} W_j \tag{7}$$

Here, $\eta_t$ & $\eta_o$, $N_t$ & $N_o$, $\alpha_t$ & $\alpha_o$ are the reference dose, number of voxels and penalty factors for target and OAR, respectively. $d_i$, $DADR_i$ are given by equations (4), (5), respectively. $D_{ij}$ is the influence matrix of dose. $I_{nozzle}$, $T_{min}$ & $N_{MU}$ are nozzle current, minimum spot duration and number of protons per MU, respectively.

**Sparse ridge filters**

Regularly spaced ridge filters, designed using the IBO-IMPT framework, provide increased DADR for some OARs while maintaining tumor coverage. However, the optimization does not take depth modulation into account. Sparse ridge filters, from which some pins are omitted, provide a means to further increase the DADR for optimal FLASH sparing.

We use a heuristic decision process to generate the sparse ridge filters. We calculate the dose influence matrices for a regular ridge filter and for a range compensator alone with no pins. The filter pin location map is used as the proton spot map, so that the dose of each beamlet reflects the contribution of a specific ridge filter pin. Using these two dose influence matrices, we obtain an optimized IBO-IMPT plan (Fig. 2A). We then derive optimized spot weighting factors, where, $w_j^r$ is the weighting factor for filter pin location $j$ of regular ridge filter and $w_j^t$ is the weighting factor for pin location $j$ of filter compensator. We then test the effect of removing the pin at position $j$ to increase DADR. We keep the pin at location $j$ if $\frac{w_j^r}{w_j^r + w_j^t} > f_j$ where $f_j$ is a user-defined threshold; otherwise, otherwise we remove the pin. The decision process is shown in Fig. 2B. After selecting the pin locations, we generate the sparse ridge filter design (Fig. 2C). The sparse filter design allows higher DADRs for OARs, including lung and heart (Fig. 2D).

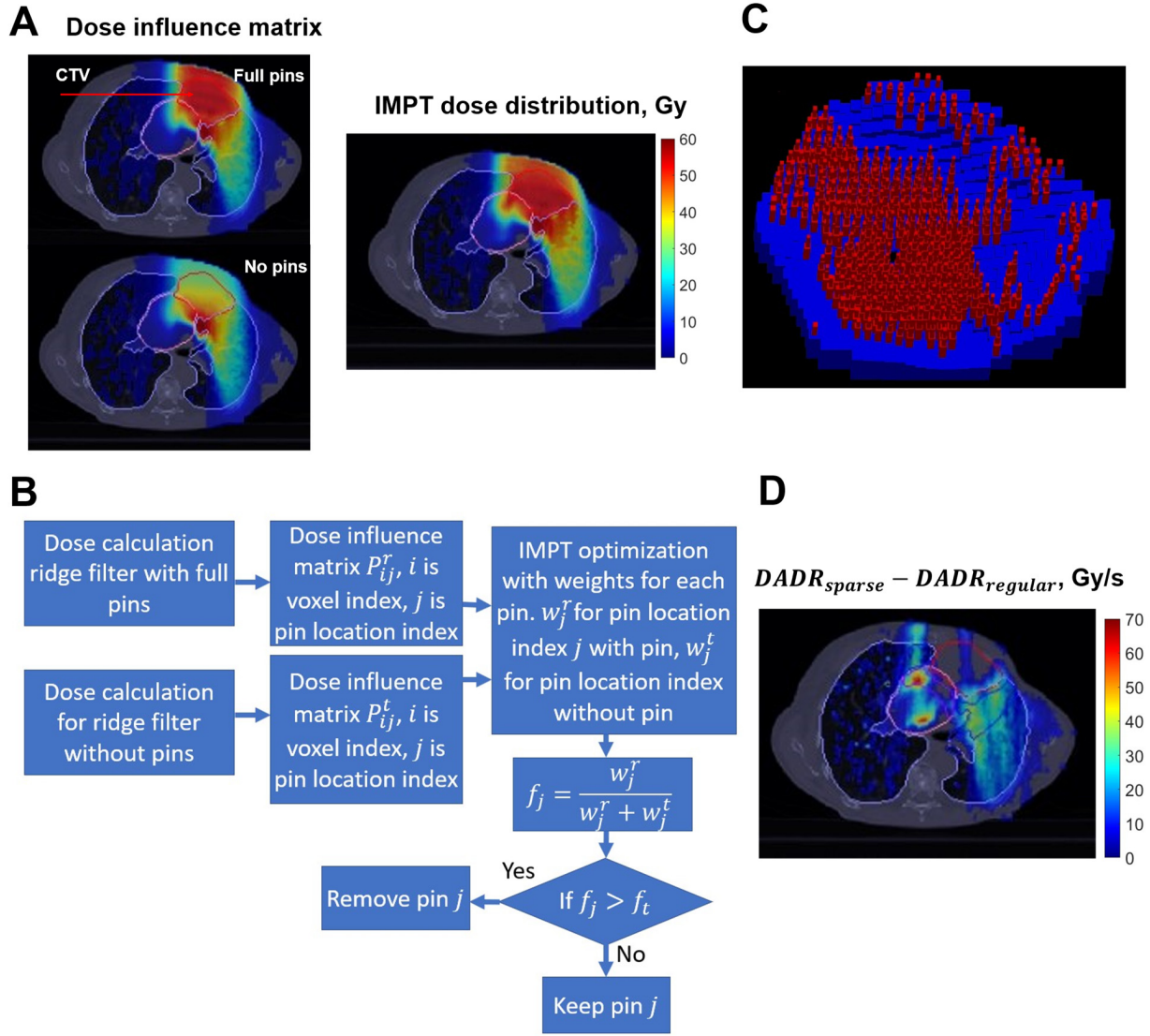

Figure 2: A. The IMPT dose distribution by two dose influence matrices by ridge filter with full pins and the ridge filter without pins. B. Heuristic decision process of generating sparse filter. C. Optimized sparse ridge filter. D. Distribution of DADR difference (DADR by sparse ridge filter minus DADR by regular ridge filter).

## Example filter design and treatment plan

To demonstrate the IBO-IMPT framework, we designed ridge filters and developed a treatment plan for an example lung cancer patient. Sparing of OARs was of particular concern, as the patient had previously been treated using conventional spot scanning proton therapy. The patient-specific ridge filter and range shifter assembly were designed to achieve conformal target dose coverage using a 250 MeV proton beam. BSPTV with 5% range uncertainty and 5 mm setup uncertainty. For our scanning beam proton therapy system, using a minimum duration of 1 millisecond and a constant current 300 nA, a value of 300 was taken as the minimum MU[37]. Three beam angles of 0º, 40º, and 320º were considered (denoted as T0, T40, and T320, respectively. The target was the clinical target volume (CTV), and two OARs, lung and heart, were considered for plan optimization. The prescribed dose to the CTV was 50 Gy (10 Gy × 5 fractions). The entire volume received a dose ≥ 50 Gy, with a maximum allowable dose for hotspots corresponding to 125% of the prescription dose (62.5 Gy). The dose volume constraints for lung and heart are shown in Table S1 in the Supplementary Materials.

IBO-IMPT plans were generated for regular and sparse ridge filters at different beam angles and compared with each other and with conventional IMPT plans as detailed in the Result section. Evaluation was performed using an approach similar to that described by Feng et al. for LET-guided optimization[38]. To generate the evaluating structures, Heart_eva, and Lung_eva, we first created a uniform 5 mm expansion of the BSPTV. The 5 mm BSPTV expansion was chosen so as to include the gradual dose fall off beyond the BSPTV, recognizing that the dose within this margin region may exceed the lower threshold for a FLASH effect[39–42]. We next removed the CTV from the expanded BSPTV and defined Lung_EVA as the overlap between this and the lung (Fig. 3). We generated the Heart_eva structure using a similar approach.

The rationale for using the only the defined Heart_eva and Lung_eva volumes, rather than the whole heart and lung, was that evaluation of a very large structure might mask the significance of high dose or high dose rate due to a large volume with a low dose and low dose rate. Fig. 3 shows the evaluation structures for the three beams. For the multiple beam plan, the overall evaluating structure is the Boolean union of the evaluation structures for each beam.

For each plan, we calculated the distribution of dose, DADR, and $LET_d$ and generated corresponding volume histograms. The FLASH effect has been reported to have a dose threshold between 4 Gy to 10 Gy [39–42]. Here, we used 4 Gy per fraction per field as a conservative estimate. The FLASH dose rate threshold has been reported to be between 40 and 100 Gy/s. Here, we used 40 Gy/s. For generating the dose rate volume histograms, we assigned the DADR as zero for the voxels that do not meet the dose threshold. Thus, we can directly observe the fraction of volume achieving FLASH by inspection of the DADR volume histogram, as only the voxels that meet the dose and dose rate thresholds contribute to the histogram.

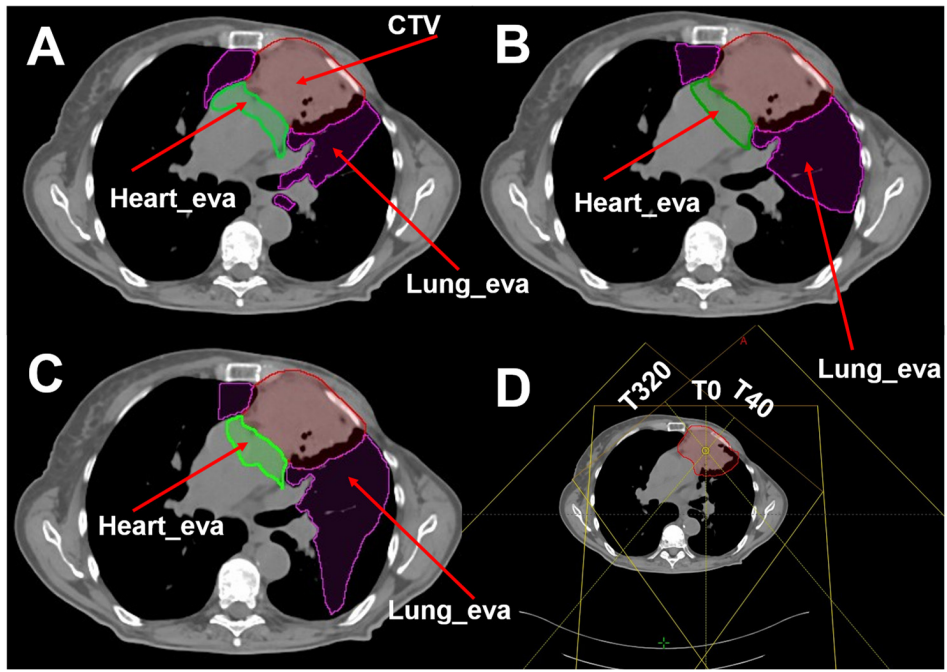

Figure 3: A. Evaluating structures. A. Heart_eva and Lung_eva for beam T40. B. Heart_eva and Lung_eva for beam T320. C. Heart_eva and Lung_eva for beam T0. D. Beam arrangement.

# Results

## IBO-IMPT with regular ridge filters

To demonstrate the functionality of the IBO-IMPT framework, we used it to design a regular ridge filter and develop a treatment plan for a sample lung cancer patient. Fig. 4 shows dose, DADR, and LETd distributions, together with volume histograms, for a plan that was developed with the goal of reducing $LET_d$ to heart while maintaining target coverage. We compare this with a conventional IMPT plan (dose optimization only). The target coverages for the IBO-IMPT and IMPT plans are similar. However, the IBO-IMPT framework resulted in a marked reduction of $LET_d$ to the heart (compare Fig. 4C and 4F, see also Fig. 4I ).

We also generated plans for several different beam orientations (T0, T40, and T320) where DADR optimization was prioritized, while maintaining adequate dose and $LET_d$ optimization (Fig. S2, S3, and S4). Together, these results demonstrate that adoption of the IBO-IMPT framework, in combination with regular ridge filters, results in at least modest improvements to DADR and $LET_d$ for OARs, while maintaining tumor coverage and meeting other constraints.

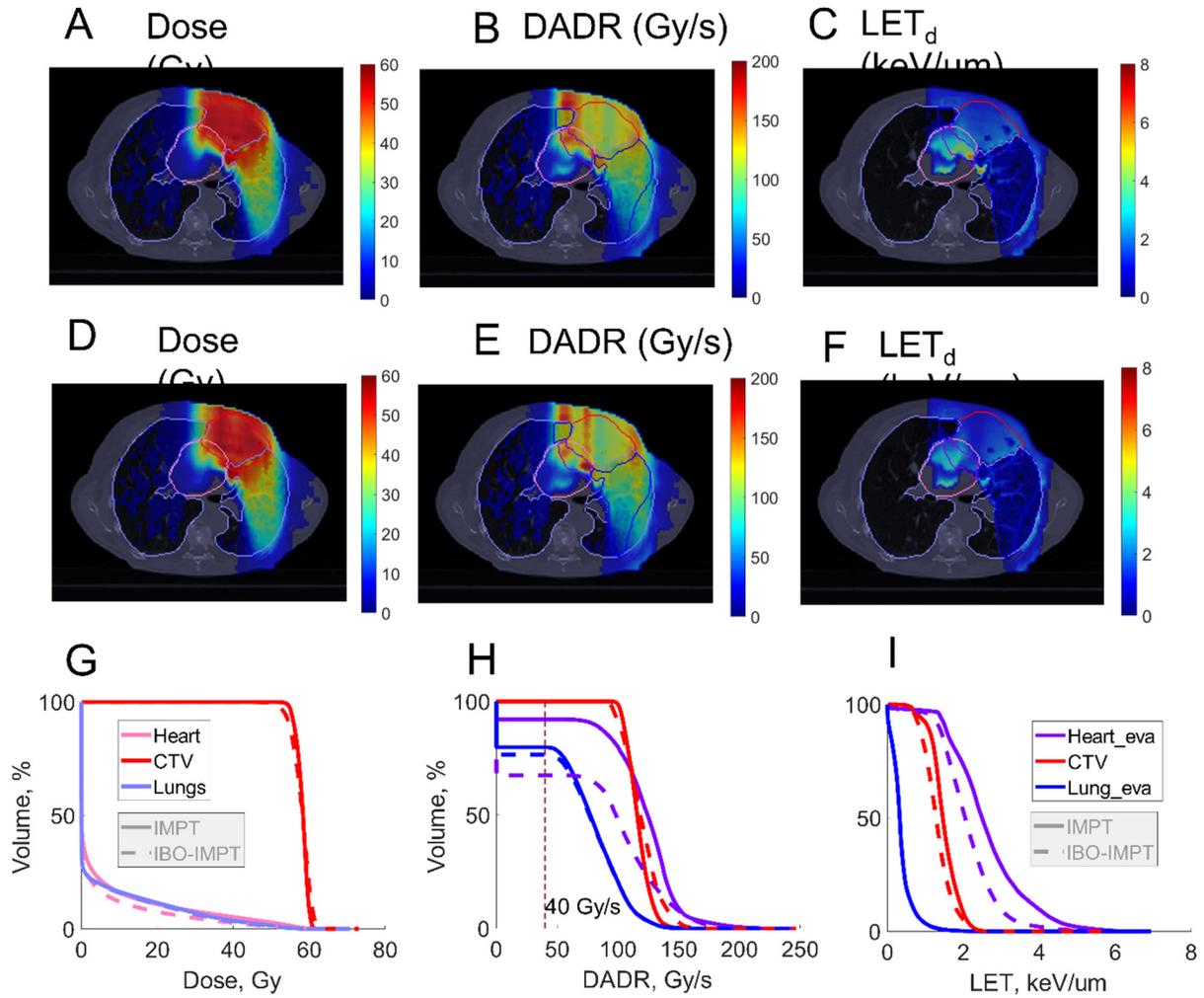

Figure 4: Treatment plans using a regular ridge filter and beam T0. A, B, C: Dose, DADR, and $LET_d$ distributions for IMPT. D, E, F: Dose, DADR, and $LET_d$ distributions for IBO-IMPT. G, H, I: Dose, DADR, $LET_d$ volume histograms. Solid lines denote the IMPT plan, and dashed line denote the IBO-IMPT plan.

## IBO-IMPT with sparse ridge filters

Regular ridge filters, originally designed for dose optimization only, cannot fully realize the benefits of the IBO-IMPT framework, because spot-specific dose-depth modulation is not optimized. To address this, we explored sparse ridge filter designs, in which some pins are removed using the heuristic decision process described in Materials and Methods.

An example of a fully optimized IBO-IMPT plan, with sparse ridge filters and multiple beams is shown in Figure 5. An IMPT-optimized plan using regular ridge filters is shown for comparison. Tumor coverage is maintained and hotspots are well controlled with both plans (compare Fig. 5A and 5D). The IBO-IMPT plan with sparse ridge filters results in a marked improvement to DADR in the OARs (compare Fig. 5B and 5E). The volume that received a dose rate of ≥40 Gy/second increased by 31% for Heart_eva and by 50% for Lung_eva increases by 50% (Fig. 5H). The $LET_d$ for the two plans was substantially the same (compare Fig. 5C and 5F, see also Fig. 5I). Together, results show that the use of sparse ridge filters and multiple beams help realize the full potential of the IBO-IMPT framework.

A separate set of optimized single-beam plans, using sparse ridge filters, is shown in Fig. S5, S6, and S7. The increased DADR for lungs, using sparse ridge filters versus regular ridge filters, is evident. The individual plans have some hotspots within BSPTV (which slightly exceed the 125% prescription dose), but sequential delivery as SBRT fractions reduces these and improves target coverage. Together, dose coverage is similar to the multi-field plan in Fig. 5, but with better FLASH sparing due to the dose threshold per fraction per field.

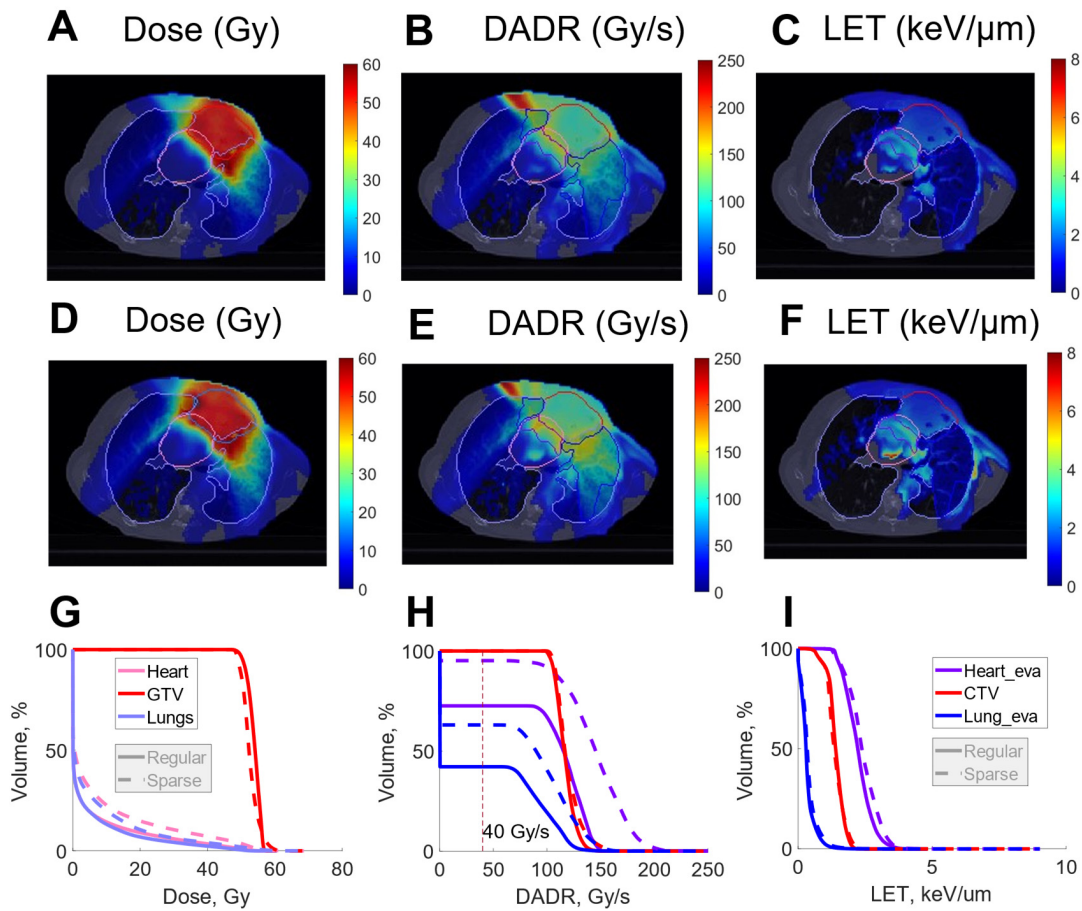

Figure 5: IMPT treatment plans with multiple beams (T0, T40, and T320). A, B, C: dose, DADR, and LET distribution of the three beams with regular ridge filter, respectively. D, E, F: dose, DADR, and LET distribution of the corresponding IBO-IMPT plans with sparse ridge filters, respectively. G, H, I: dose, DADR, LET volume histograms, respectively. The solid line for the plan with regular ridge filter, and dash line for the plan with sparse ridge filter.

**Preliminary dose verification with a patient-specific ridge filter**

To verify the ability of the ridge filter assembly to deliver the predicted dose, we performed proton dose measurements. The ridge filter assembly (shown in Fig 1D), which consists of filter pins and a compensator, was placed on T0 beam axis (Fig. S8A). A range shifter, solid water, and an ionization chamber array were placed downstream (Fig. S8B). We delivered a treatment plan optimized for our example lung cancer patient and designed to provide uniform dose to CTV target. The calculated dose distribution, 25 mm depth from the solid water surface is shown in Fig. S8C. The corresponding measured dose is shown in Fig. S8D. The total gamma passing rate was 92% (3mm/3%) (Fig. S8), which exceeds the standard patient QA passing criteria of 90%. Results provide a preliminary demonstration that the ridge filter assembly can facilitate delivery of a clinically acceptable dose distribution. Further verification of FLASH timing, dose rate, and LET, using a novel time-resolved and spatially-resolved detector, is in progress and will be reported separately.

## Discussion

In this feasibility study, we demonstrate how the use of the IBO-IMPT framework and optimized ridge filters can improve DADR and $LET_d$, for lung and heart, relative to a plan generated using a standard IMPT approach. The IBO-IMPT framework, which explicitly incorporates objective functions of dose, DADR, and $LET_d$, provides degenerate solutions for patient-specific ridge filter and spot maps while providing an ability to study the contribution of each term[27]. Optimization of DADR and $LET_d$, while maintaining a similar dose distribution, is crucial for disentangling the biological contributions of DADR and LET from that of dose *per se*. For OARs such as great vessels, which have a maximum tolerated dose close to the prescription dose, increasing the DADR above the FLASH threshold (≥40 Gy/sec) may be the best selection. Alternatively, lower dose or LET might be a better option for OARs, such as spinal cord, which have a maximum tolerated dose that is smaller than the prescription dose. Such options can be explored using the IBO-IMPT optimization framework.

One of applications of IBO-IMPT is the design of sparse ridge filter/compensator assemblies, from which some pins have been omitted. The sparse ridge filters are more efficient than regular filters, providing more flexibility to improve the DADR. Use of the sparse filters can lead to some hotspots within the CTV, although this can be mitigated by alternating the beam orientation over SBRT fractions. The sparse ridge filter design process is currently based on a heuristic method. Several trial and error iterations are generally required to achieve an acceptable result. It should be possible to improve on this approach, for example by developing a faster dose calculation engine for patient-specific ridge filters, which would allow combination of the ridge filter and plan optimization processes through a stepwise optimization scheme or using mixed-integer programming. This would allow simultaneous optimization of the proton spot map and the filter pin location map.

It is important to note that the biological mechanism of FLASH sparing remains a subject of active investigation. IBO-IMPT optimization can assist this work by enabling biologists to separate the contribution of LET from dose rate effects. With the IBO-IMPT framework, different beam designs can be examined in parallel to determine the contribution of each term. When better biological models of the FLASH effect are available, the IBO-IMPT can be extended to incorporate them directly, rather than indirectly via DADR and $LET_d$ terms. Other examples include replacing the DADR with other dose rate approaches in IBO-IMPT[16,43]. The work here assumed a constant beam current, which allowed a simplified optimization model for DADR, keeping spot MUs as the sole decision variables. Solutions for adding current as a decision variable exist in literature[44], however, and could be integrated into the IBO-IMPT framework in the future.

## Conclusion

This proof-of-concept study demonstrates to solve for dose, DADR, and LETd, using an integrated biological optimization framework. Example solutions include regular and sparse patient-specific ridge filters and spot maps designed prioritize dose, DADR, and LETd for optimal sparing of heart and lung. This novel method will facilitate delivery of conformal proton fields at FLASH rates for preclinical and clinical studies.

## Acknowledgment

We thank Dr. Hao Gao (University of Kansas Medical Center) for helpful discussions of about the applicability of simultaneous optimization of dose and dose rate (SDDR) and minimum MU constraint to ridge optimization. We thank Dr. Yuning Hou and the Cancer Animal Models shared resource of the Winship Cancer Institute of Emory University (supported by NIH/NCI award number P30CA138292) for micro-CT imaging and helpful discussions. We thank Nathan Harrison (Emory Proton Therapy Center) for editing and discussing the manuscript. Work was supported by Emory University startup funding (LL, RL), by the German Research Foundation (DFG) grant number  WA 4707/1-1 (NW), by the National Aeronautics and Space Administration (award number 80NSSC18K1116 to WSD), by Vanderbilt University (subcontract under NIH/NCI grant 1R21CA226562).

# Reference

1. Darby, S. C. *et al.* Risk of Ischemic Heart Disease in Women after Radiotherapy for Breast Cancer. *N. Engl. J. Med.* **368**, 987–998 (2013).

2. Bradley, J. D. *et al.* Standard-dose versus high-dose conformal radiotherapy with concurrent and consolidation carboplatin plus paclitaxel with or without cetuximab for patients with stage IIIA or IIIB non-small-cell lung cancer (RTOG 0617): A randomised, two-by-two factorial p. *Lancet Oncol.* **16**, 187–199 (2015).

3. Modh, A. *et al.* Local control and toxicity in a large cohort of central lung tumors treated with stereotactic body radiation therapy. *Int. J. Radiat. Oncol. Biol. Phys.* **90**, 1168–1176 (2014).

4. Stam, B. *et al.* Dose to heart substructures is associated with non-cancer death after SBRT in stage I–II NSCLC patients. *Radiother. Oncol.* **123**, 370–375 (2017).

5. Chi, A., Chen, H., Wen, S., Yan, H. & Liao, Z. Comparison of particle beam therapy and stereotactic body radiotherapy for early stage non-small cell lung cancer: A systematic review and hypothesis-generating meta-analysis. *Radiother. Oncol.* **123**, 346–354 (2017).

6. Chang, J. Y. *et al.* Consensus Guidelines for Implementing Pencil-Beam Scanning Proton Therapy for Thoracic Malignancies on Behalf of the PTCOG Thoracic and Lymphoma Subcommittee. *Int. J. Radiat. Oncol. Biol. Phys.* **99**, 41–50 (2017).

7. Chang, J. Y. *et al.* Consensus Statement on Proton Therapy in Early-Stage and Locally Advanced Non-Small Cell Lung Cancer. *Int. J. Radiat. Oncol. Biol. Phys.* **95**, 505–516 (2016).

8. Lin, L. *et al.* Beam-specific planning target volumes incorporating 4D CT for pencil beam scanning proton therapy of thoracic tumors. *J. Appl. Clin. Med. Phys.* **16**, 281–292 (2015).

9. Lin, L. *et al.* Evaluation of motion mitigation using abdominal compression in the clinical implementation of pencil beam scanning proton therapy of liver tumors: *Med. Phys.* **44**, 703–712 (2017).

10. Matney, J. *et al.* Effects of respiratory motion on passively scattered proton therapy versus

intensity modulated photon therapy for stage III lung cancer: Are proton plans more sensitive to breathing motion? *Int. J. Radiat. Oncol. Biol. Phys.* **87**, 576–582 (2013).

11. Liu, W. *et al.* Impact of respiratory motion on worst-case scenario optimized intensity modulated proton therapy for lung cancers. *Pract. Radiat. Oncol.* **5**, e77–e86 (2015).

12. Tryggestad, E. J., Liu, W., Pepin, M. D., Hallemeier, C. L. & Sio, T. T. Managing treatment-related uncertainties in proton beam radiotherapy for gastrointestinal cancers. *J. Gastrointest. Oncol.* **11**, 212–224 (2020).

13. Nantavithya, C. *et al.* Phase 2 Study of Stereotactic Body Radiation Therapy and Stereotactic Body Proton Therapy for High-Risk, Medically Inoperable, Early-Stage Non-Small Cell Lung Cancer. *Int. J. Radiat. Oncol. Biol. Phys.* **101**, 558–563 (2018).

14. Levy, K. *et al.* Abdominal FLASH irradiation reduces radiation-induced gastrointestinal toxicity for the treatment of ovarian cancer in mice. *Sci. Rep.* **10**, 1–14 (2020).

15. Diffenderfer, E. S. *et al.* Design, Implementation, and in Vivo Validation of a Novel Proton FLASH Radiation Therapy System. *Int. J. Radiat. Oncol. Biol. Phys.* **106**, 440–448 (2020).

16. Kang, M., Wei, S., Isabelle Choi, J., Simone, C. B. & Lin, H. Quantitative assessment of 3D dose rate for proton pencil beam scanning FLASH radiotherapy and its application for lung hypofractionation treatment planning. *Cancers (Basel).* **13**, 1–14 (2021).

17. Gerbershagen, A. *et al.* Measurements and simulations of boron carbide as degrader material for proton therapy. *Phys. Med. Biol.* **61**, N337–N348 (2016).

18. Simeonov, Y. *et al.* 3D range-modulator for scanned particle therapy: Development, Monte Carlo simulations and experimental evaluation. *Phys. Med. Biol.* **62**, 7075–7096 (2017).

19. Brouwer, L., Huggins, A. & Wan, W. An achromatic gantry for proton therapy with fixed-field superconducting magnets. *Int. J. Mod. Phys. A* **34**, (2019).

20. Diffenderfer, E. S. *et al.* Design, Implementation, and in Vivo Validation of a Novel Proton FLASH Radiation Therapy System. *Int. J. Radiat. Oncol. Biol. Phys.* **106**, 440–448

(2020).

21. Zou, W. *et al.* Characterization of a high-resolution 2D transmission ion chamber for independent validation of proton pencil beam scanning of conventional and FLASH dose delivery. *Med. Phys.* **48**, 3948–3957 (2021).

22. Kang, M., Wei, S., Choi, J. I., Lin, H. & Simone, C. B. A Universal Range Shifter and Range Compensator Can Enable Proton Pencil Beam Scanning Single-Energy Bragg Peak FLASH-RT Treatment Using Current Commercially Available Proton Systems. *Int. J. Radiat. Oncol.* 1–11 (2022) doi:10.1016/j.ijrobp.2022.01.009.

23. Wei, S., Lin, H., Choi, J. I., Simone, C. B. & Kang, M. A novel proton pencil beam scanning flash rt delivery method enables optimal oar sparing and ultra-high dose rate delivery: A comprehensive dosimetry study for lung tumors. *Cancers (Basel).* **13**, (2021).

24. Kim, M. M. *et al.* Comparison of flash proton entrance and the spread-out bragg peak dose regions in the sparing of mouse intestinal crypts and in a pancreatic tumor model. *Cancers (Basel).* **13**, (2021).

25. Evans, T., Cooley, J., Wagner, M., Yu, T. & Zwart, T. Demonstration of the FLASH Effect Within the Spread-out Bragg Peak After Abdominal Irradiation of Mice. *Int. J. Part. Ther.* **8**, 68–75 (2022).

26. Zhang, G., Gao, W. & Peng, H. Design of static and dynamic ridge filters for FLASH–IMPT: A simulation study. *Med. Phys.* 1–13 (2022) doi:10.1002/mp.15717.

27. van de Water, S., Safai, S., Schippers, J. M., Weber, D. C. & Lomax, A. J. Towards FLASH proton therapy: the impact of treatment planning and machine characteristics on achievable dose rates. *Acta Oncol. (Madr).* **58**, 1463–1469 (2019).

28. Deng, W. *et al.* Hybrid 3D analytical linear energy transfer calculation algorithm based on precalculated data from Monte Carlo simulations. *Med. Phys.* **47**, 745–752 (2020).

29. Deng, W. *et al.* A critical review of LET-based intensity- modulated proton therapy plan evaluation and optimization for head and neck cancer management. *Int. J. Part. Ther.* **8**, 36–49 (2021).

30. Park, P. C. *et al.* A beam-specific planning target volume (PTV) design for proton therapy to account for setup and range uncertainties. *Int. J. Radiat. Oncol. Biol. Phys.* **82**, e329–e336 (2012).

31. 3D Systems. Accura Xtreme Datasheet. *https://www.3dsystems.com/materials/accura-xtreme/tech-specs* (2022).

32. Shan, J., An, Y., Bues, M., Schild, S. E. & Liu, W. Robust optimization in IMPT using quadratic objective functions to account for the minimum MU constraint. *Med. Phys.* **45**, 460–469 (2018).

33. An, Y. *et al.* Robust intensity-modulated proton therapy to reduce high linear energy transfer in organs at risk. *Med. Phys.* **44**, 6138–6147 (2017).

34. Wieser, H.-P. *et al.* Development of the open-source dose calculation and optimization toolkit matRad. *Med. Phys* **44**, (2556).

35. Bangert, M. *et al.* e0404/matRad: Blaise v2.10.0. (2020) doi:10.5281/ZENODO.3879616.

36. Wächter, A. & Biegler, L. T. On the implementation of an interior-point filter line-search algorithm for large-scale nonlinear programming. *Math. Program.* **106**, 25–57 (2006).

37. Wei, S. *et al.* FLASH Radiotherapy Using Single-Energy Proton PBS Transmission Beams for Hypofractionation Liver Cancer: Dose and Dose Rate Quantification. *Front. Oncol.* **11**, 1–11 (2022).

38. Feng, H. *et al.* Per-voxel constraints to minimize hot spots in linear energy transfer-guided robust optimization for base of skull head and neck cancer patients in IMPT. *Med. Phys.* **49**, 632–647 (2022).

39. Chabi, S. *et al.* Ultra-high-dose-rate FLASH and Conventional-Dose-Rate Irradiation Differentially Affect Human Acute Lymphoblastic Leukemia and Normal Hematopoiesis. *Int. J. Radiat. Oncol. Biol. Phys.* **109**, 819–829 (2021).

40. Montay-Gruel, P. *et al.* Hypofractionated FLASH-RT as an effective treatment against glioblastoma that reduces neurocognitive side effects in mice. *Clin. Cancer Res.* **27**, 775–784 (2021).

41. Wilson, J. D., Hammond, E. M., Higgins, G. S. & Petersson, K. Ultra-High Dose Rate (FLASH) Radiotherapy: Silver Bullet or Fool's Gold? *Front. Oncol.* **9**, 1–12 (2020).

42. Gao, H. *et al.* Simultaneous dose and dose rate optimization (SDDRO) of the FLASH effect for pencil-beam-scanning proton therapy. *Med. Phys.* **49**, 2014–2025 (2022).

43. Folkerts, M. M. *et al.* A framework for defining FLASH dose rate for pencil beam scanning. *Med. Phys.* **47**, 6396–6404 (2020).

44. Gao, H. *et al.* Simultaneous dose and dose rate optimization (SDDRO) for FLASH proton therapy. *Med. Phys.* **47**, 6388–6395 (2020).
22

# Supplementary Materials

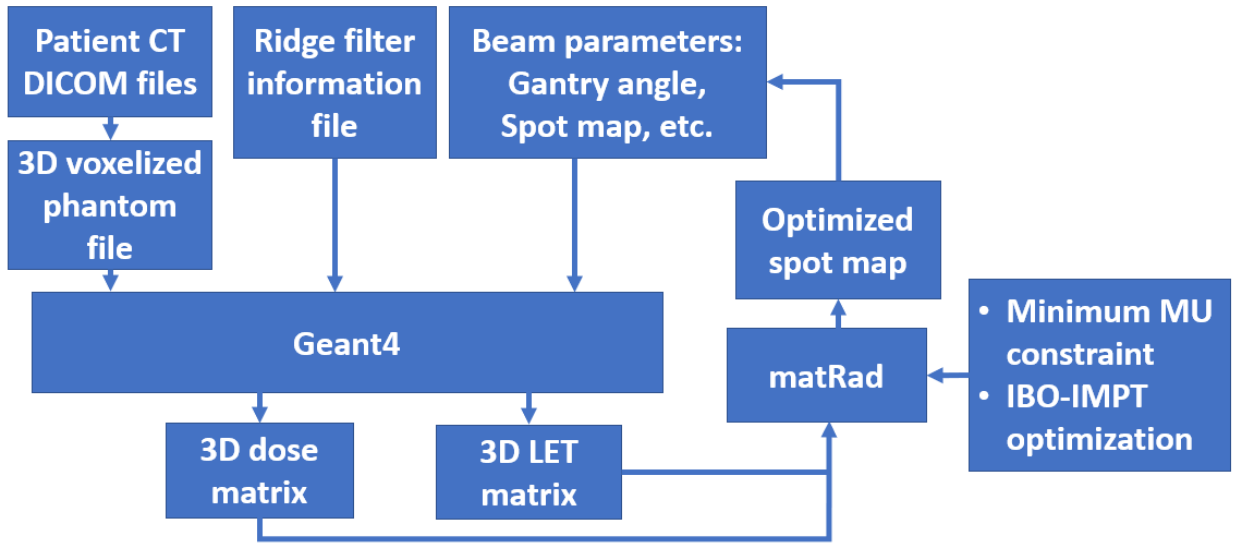

Figure S1: Flow chart of designed in-house TPS

Table S1: Treatment planning OAR optimization constraints

| Organ at risk | constraint |
|---|---|
| Heart | $D_{mean} \leq 20$ Gy |
| Heart | V30Gy≤50% |
| Heart | $D_{max}$ = 52.5 Gy |
| Lungs | V20Gy≤35% |
| Lungs | V5Gy≤60% |
| Lungs | $D_{mean} \leq 20$ Gy |

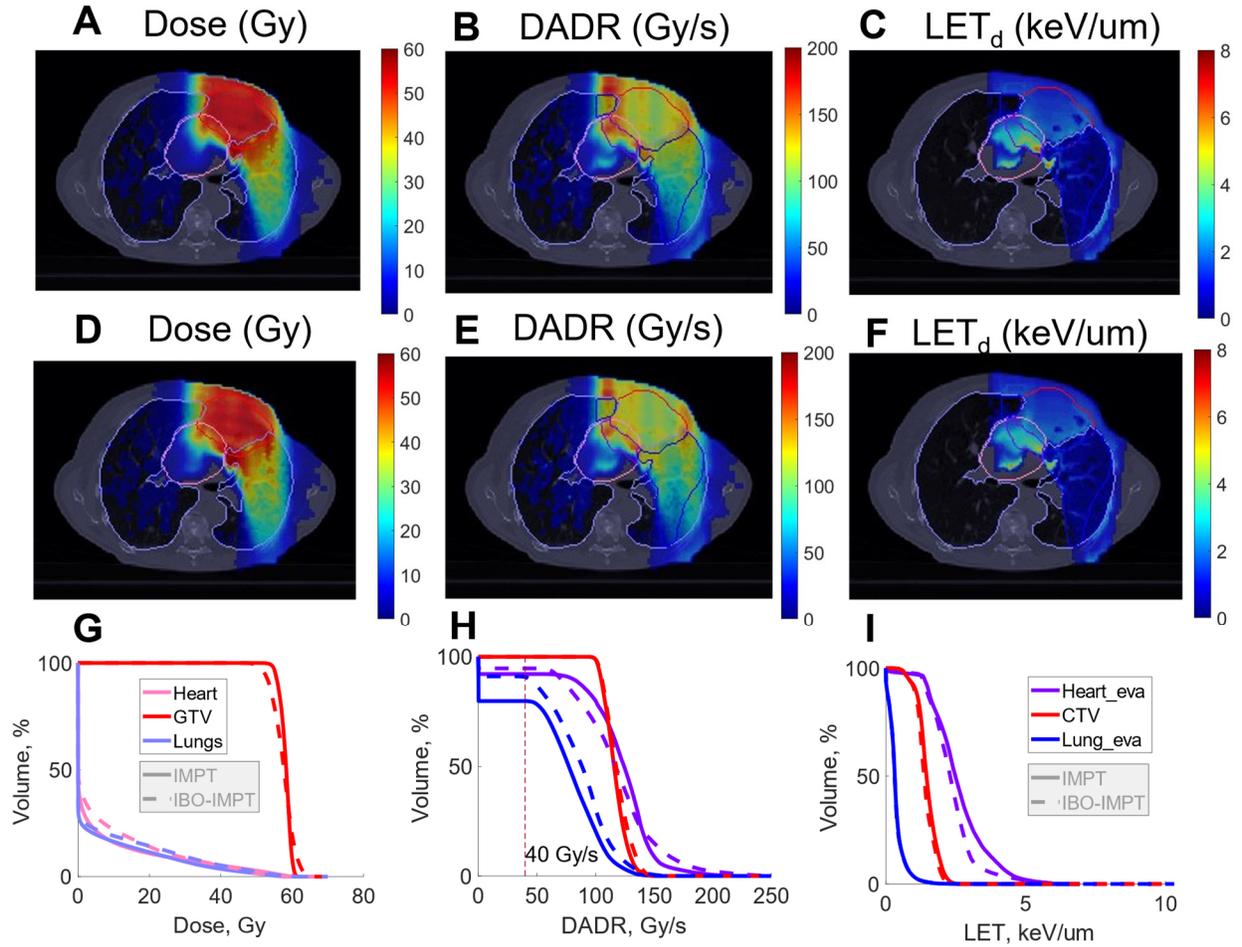

Figure S2: Treatment plans with beam T0. A, B, C: dose, DADR, and LET distribution for regular ridge filter with IMPT, respectively. D, E, F: dose, DADR, LET distribution for regular ridge filter with IBO-IMPT, respectively. G, H, I: dose, DADR, LET volume histograms, respectively. Solid line for IMPT plan, dash line for IBO-IMPT plan.

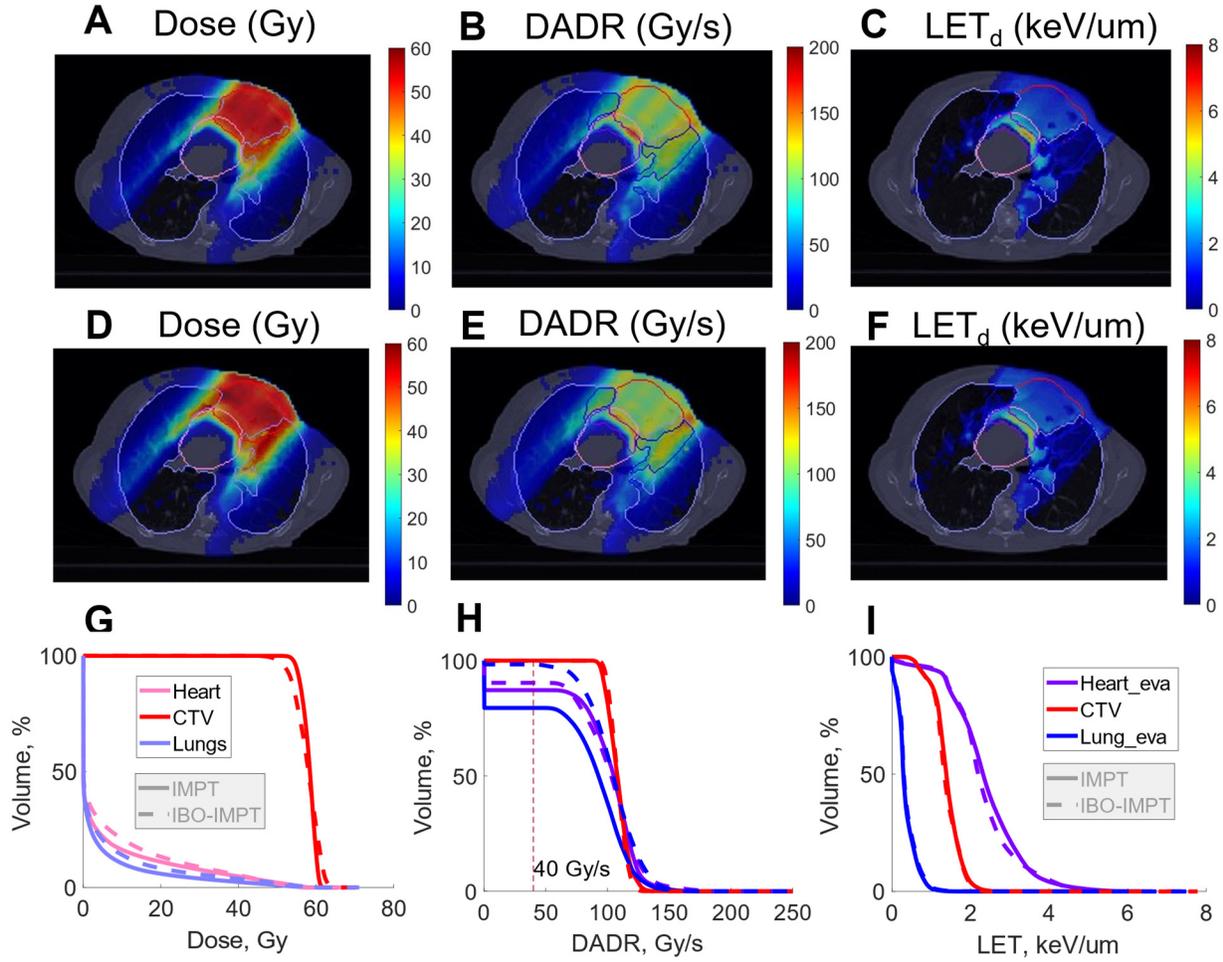

Figure S3: Treatment plans with beam T40. A, B, C: dose, DADR, and LET distribution for regular ridge filter with IMPT, respectively. D, E, F: dose, DADR, LET distribution for regular ridge filter with IBO-IMPT, respectively. G, H, I: dose, DADR, LET volume histograms, respectively. Solid line for IMPT plan, dash line for IBO-IMPT plan.

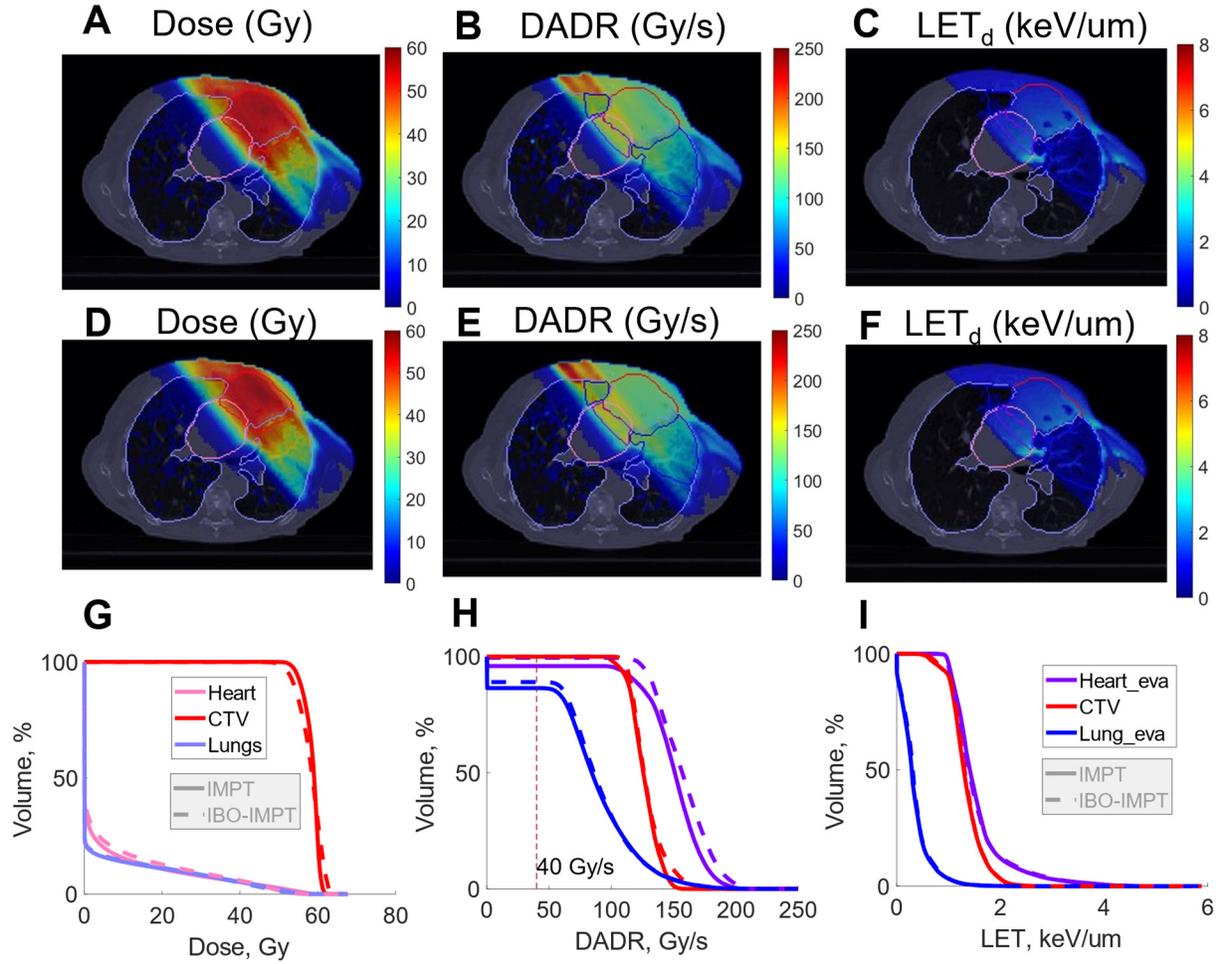

Figure S4: Treatment plans with beam T320. A, B, C: dose, DADR, and LET distribution for regular ridge filter with IMPT, respectively. D, E, F: dose, DADR, LET distribution for regular ridge filter with IBO-IMPT, respectively. G, H, I: dose, DADR, LET volume histograms, respectively. Solid line for IMPT plan, dash line for IBO-IMPT plan.

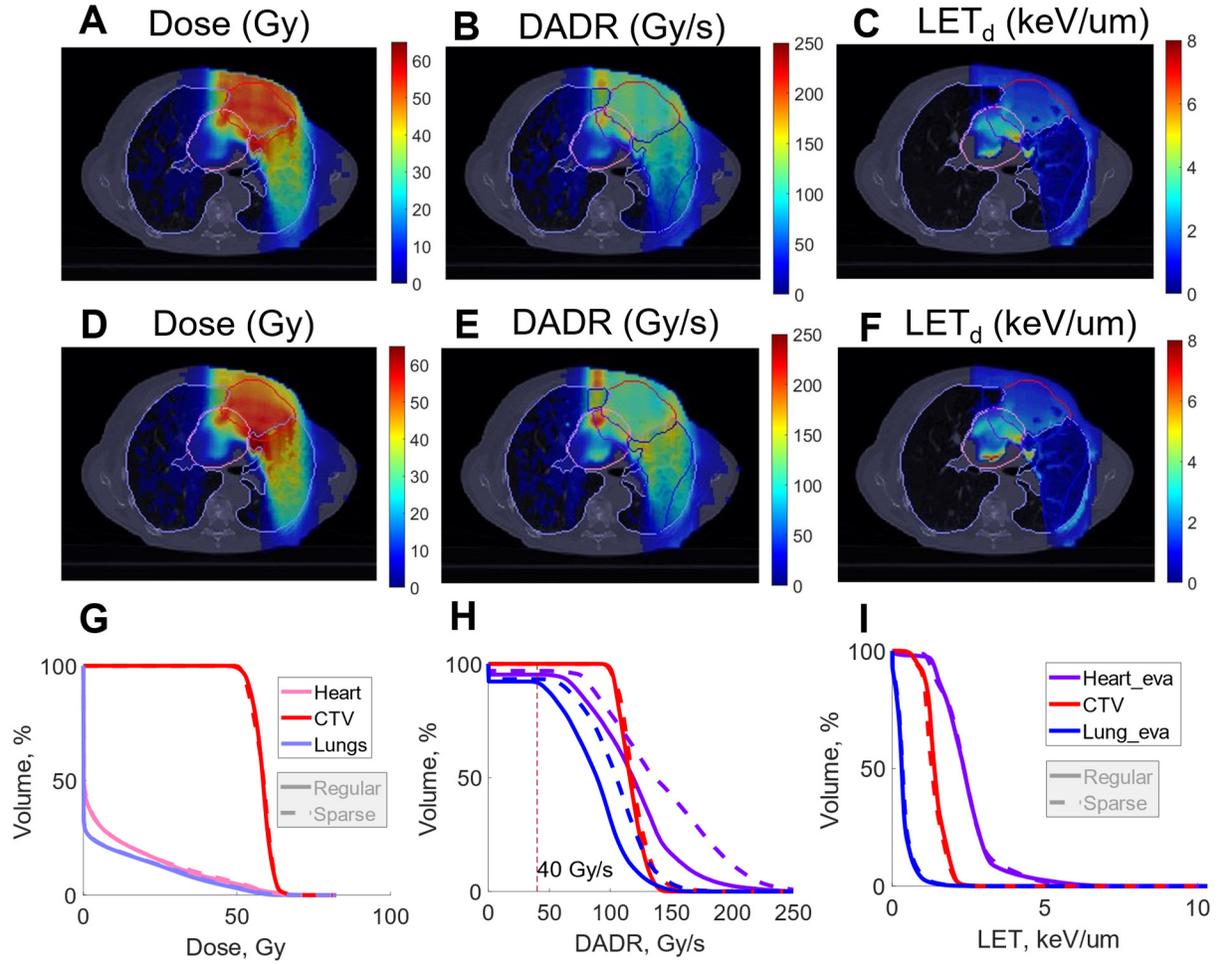

Figure S5: Treatment plans with beam T0. A, B, C: dose, DADR, and LET distribution for regular ridge filter with IBO-IMPT, respectively. D, E, F: dose, DADR, LET distribution for sparse ridge filter with IBO-IMPT, respectively. G, H, I: dose, DADR, LET volume histograms, respectively. Solid line for IMPT plan, dash line for IBO-IMPT plan. For fair comparison, the optimization constraints are the same to generate the competing plans.

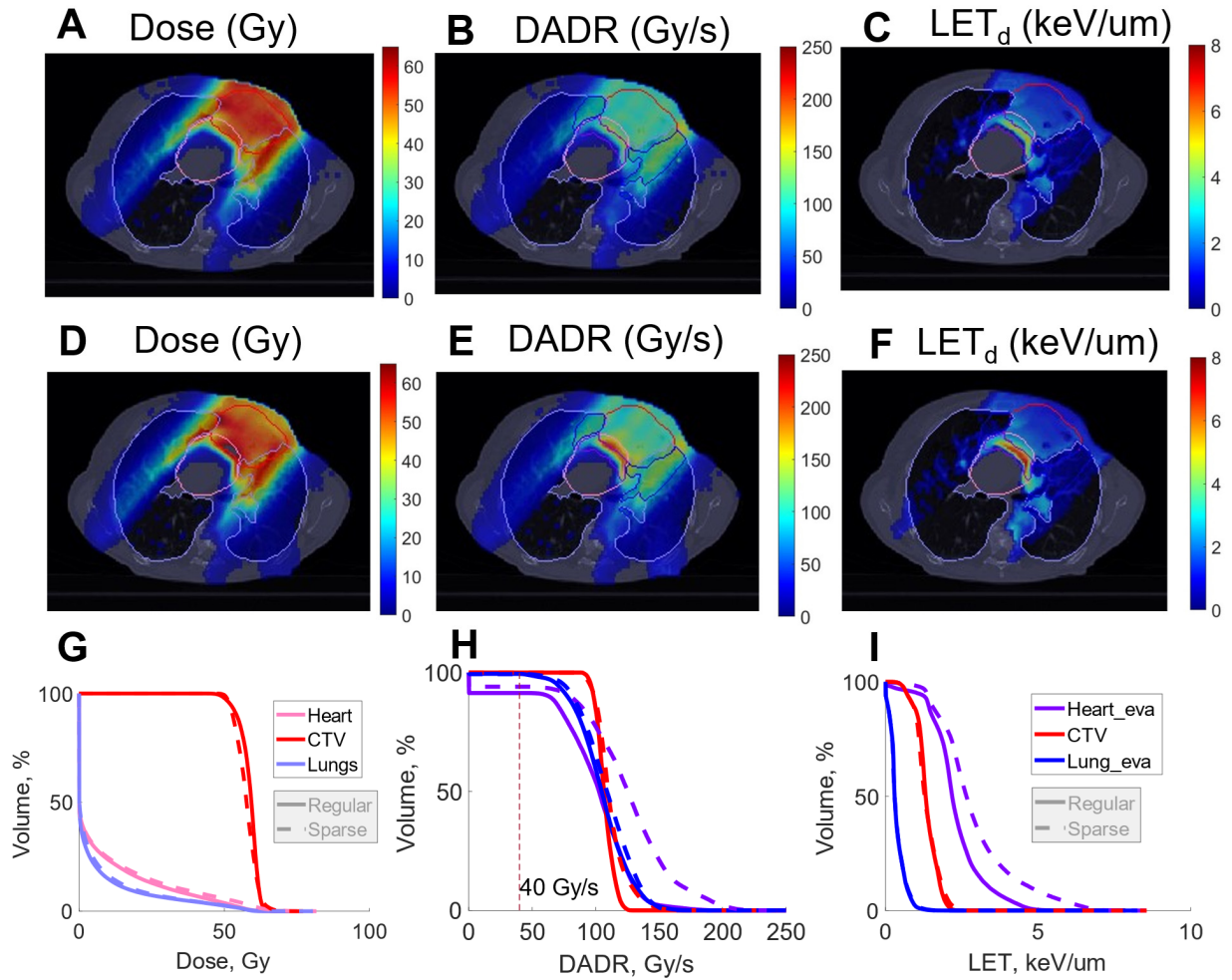

Figure S6: Treatment plans with beam T40. A, B, C: dose, DADR, and LET distribution for regular ridge filter with IBO-IMPT, respectively. D, E, F: dose, DADR, LET distribution for sparse ridge filter with IBO-IMPT, respectively. G, H, I: dose, DADR, LET volume histograms, respectively. Solid line for IMPT plan, dash line for IBO-IMPT plan. For fair comparison, the optimization constraints are the same to generate the competing plans.

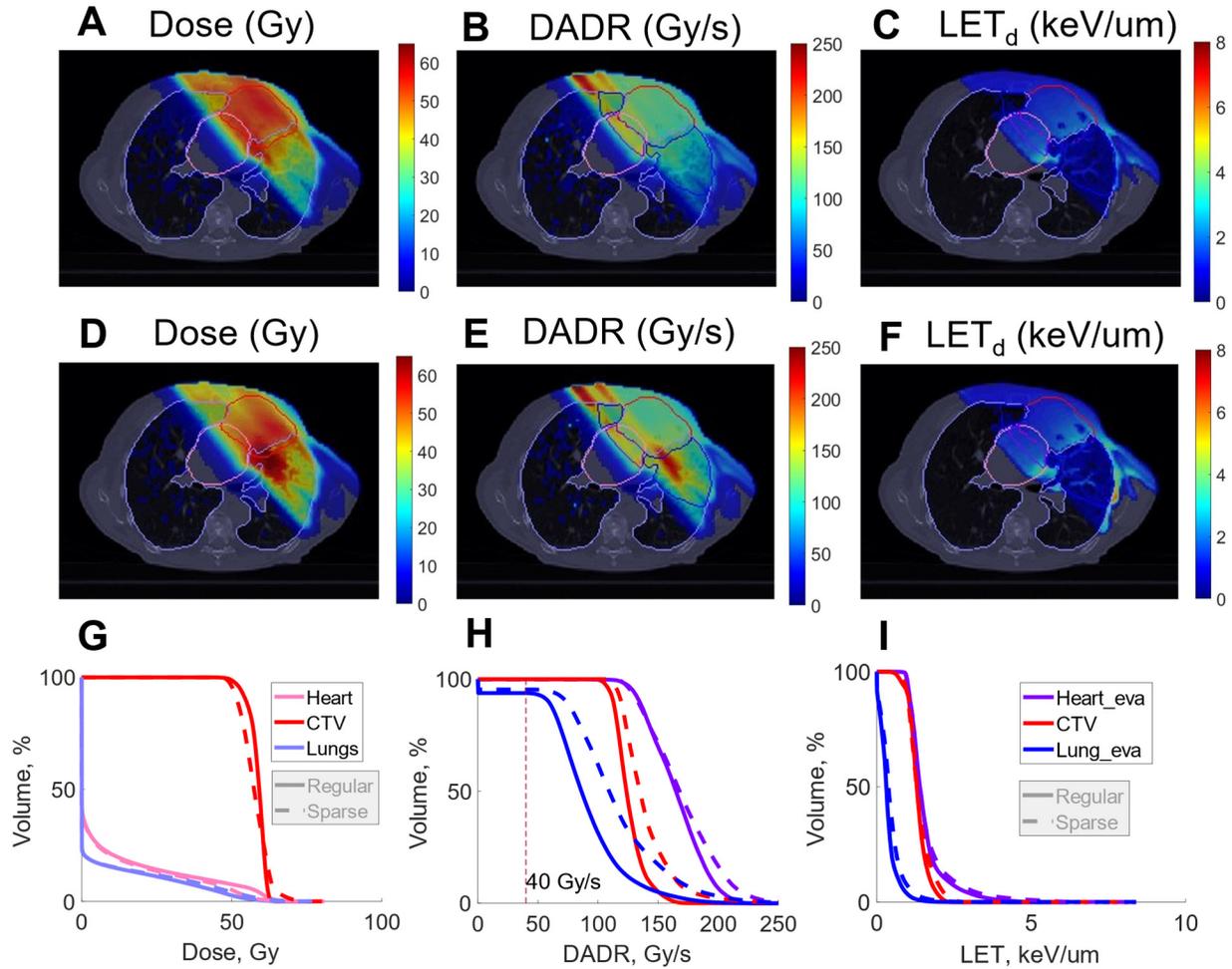

Figure S7: Treatment plans with beam T320. A, B, C: dose, DADR, and LET distribution for regular ridge filter with IBO-IMPT, respectively. D, E, F: dose, DADR, LET distribution for sparse ridge filter with IBO-IMPT, respectively. G, H, I: dose, DADR, LET volume histograms, respectively. Solid line for IMPT plan, dash line for IBO-IMPT plan. For fair comparison, the optimization constraints are the same to generate the competing plans.
29

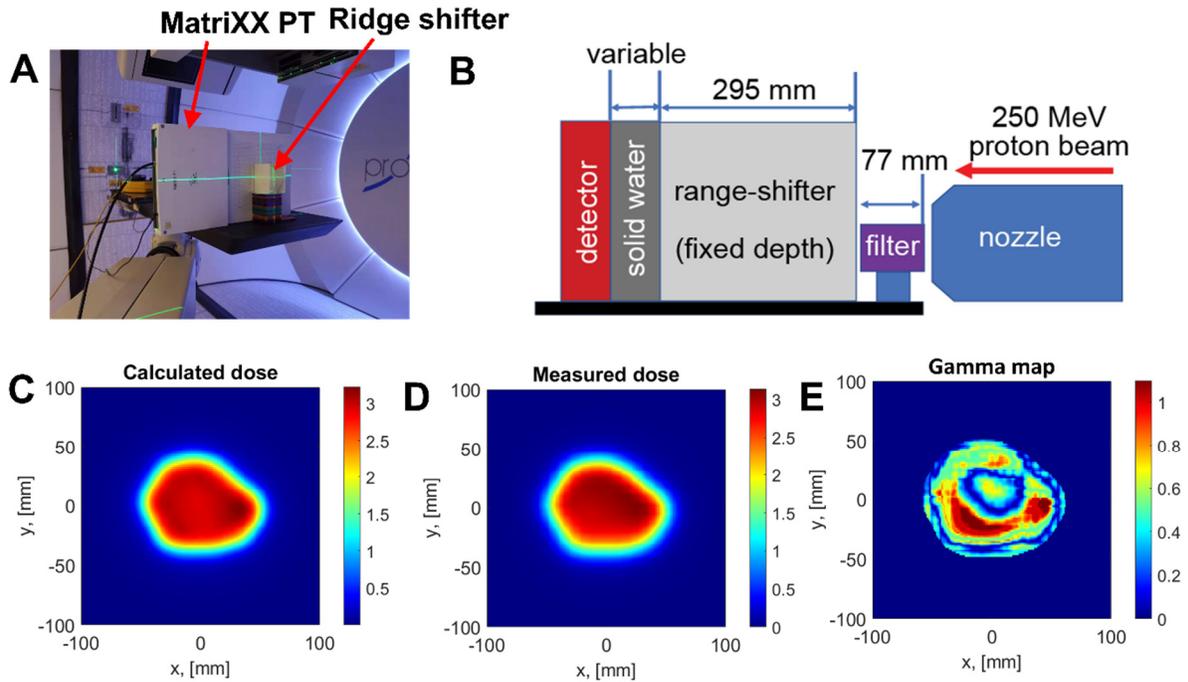

Figure S8: Dosimetry and experimental setup. A. MatriXX PT dose measurement setup for measuring the dose distribution of 250 MeV proton beam after passing the ridge filter. B. Schematic of the experimental setup displaying the beamline, ridge filter, solid water range shifter, and MatriXX PT. C. Calculated dose distribution at 25 mm depth from the solid water surface. D. Measured dose distribution at 25 mm depth from solid water surface after interpolation to the same voxel size used in dose calculation. During irradiation, the proton current was set up as 10 nA, and the minimum MU number was 30. E. Gamma map for comparing the calculated dose and the measured dose.